\documentclass[12pt]{article}

\usepackage[cp866]{inputenc}
\usepackage{graphicx}

\newcommand{\bcs}{\mbox{\scriptsize BCS}}
\newcommand{\aaa}{\mbox{\scriptsize A}}
\newcommand{\bbb}{\mbox{\scriptsize B}}
\newcommand{\fl}{\mbox{\scriptsize FL}}

\begin{document}

\newcommand{\beq}{\begin{equation}}

\newcommand{\eeq}{\end{equation}}
\begin{center}

\begin{Large}
{\bf Nodes of the Gap Function and Anomalies
in Thermodynamic Properties of Superfluid $^3$He}
\end{Large}
\\
\vskip .7truecm

\vskip 1 cm
\begin{large}
M.~V.~Zverev$^1$, V.~A.~Khodel$^{1,2}$,  J.~W.~Clark$^2$
\end{large}
\vskip 1 cm

{\small $^1$ Russian Research Centre Kurchatov Institute, Moscow,
123182, Russia}

{\small $^2$ McDonnell Center for the Space Sciences and
Department of Physics, \\ Washington University, St.~Louis, MO
63130, USA}

\date{\today}

\end{center}

\vskip 1 cm

\begin{abstract}

Departures of thermodynamic properties of three-dimensional
superfluid $^3$He from the predictions of BCS theory are analyzed.
Attention is focused on deviations of the ratios $\Delta(T=0)/T_c
$ and $\left[C_s(T_c) - C_n(T_c)\right]/C_n(T_c)$ from their
BCS values, where $\Delta(T=0)$ is the pairing gap at zero
temperature, $T_c$ is the critical temperature, and $C_s$ and
$C_n$ are the superfluid and normal specific heats.  We attribute
these deviations to the momentum dependence of the gap function
$\Delta( p)$, which becomes well pronounced when this function has
a pair of nodes lying on either side of the Fermi surface.  We
demonstrate that such a situation arises if the $P$-wave pairing
interaction ${\cal V}(p_1,p_2)$, evaluated at the Fermi surface,
has a sign opposite to that anticipated in B
CS theory.  Taking
account of the momentum structure of the gap function, we derive a
closed relation between the two ratios that contains no adjustable
parameters and agrees with the experimental data. Some important
features of the effective pairing interaction are inferred from
the analysis.
\end{abstract}

\section{Introduction}

Non-Fermi-liquid (NFL) behavior in normal states of strongly
correlated Fermi systems, as reflected in discrepancies between
experimental data and predictions of Landau's original quasiparticle
picture \cite{lan}, has been the subject of intense debate during the
past decade.  To deal with superfluid states of fermionic systems,
 the Landau picture -- involving immortal quasiparticles and no
damping -- has been joined with BCS theory to form the standard
Landau Fermi-liquid approach.   This approach is adequate for
describing conventional (``low-temperature'') superconductors.
However, significant deviations from its predictions have been
observed in experiments on strongly correlated Fermi systems.
This evidence of NFL behavior has received little attention,
despite the fact that it has been seen at extremely low temperatures
where the Landau-BCS theory should be most effective.

Conspicuous examples of NFL behavior of superfluid states are
found in three-dimen\-sional (3D) liquid $^3$He.  Specifically, in
the B-state at the melting point, the recently measured ratio
$r_\Delta = \Delta(T=0)/T_c= 1.99\pm 0.05$ \cite{todo} of the
zero-temperature gap value $\Delta(T=0)$ to the critical
temperature $T_c$ exceeds the familiar BCS value 1.76 by 13\%.
Additionally, the ratio $r_C =
\left[C_s(T_c)-C_n(T_c)\right]/C_n(T_c) $ of the normal-superfluid
jump in specific heat, evaluated at the critical point, is
significantly greater than the BCS value of 1.43. This effect is
especially prominent close to solidification: in the B-state, the
excess reaches 30\%, and it is even greater ($\sim~50\%$) in the
A-state \cite{grey}.  Such departures suggest that the source of
error lies in the conventional, oversimplified form of the BCS
pairing interaction, rather
 than in a failure of the general Landau picture.  It
 is our purpose here to explore the former, less radical
alternative.

It is an unfortunate feature of BCS theory in application to
superfluid $^3$He that the B-state always wins the energetic
competition with the A-state, whereas experimentally the A-phase
occupies a substantial portion of the phase diagram.  To address
this theoretical shortcoming, various authors
\cite{ab,ms,ba,bsa,leg,vw} have introduced modifications of the
pairing interaction ${\cal V}$ (``strong-coupling corrections'')
inherent to the presence of a pair condensate, noting that in BCS
theory ${\cal V}$ is evaluated in the normal state.  The original
idea, pioneered by Anderson and Brinkman in Ref.~\cite{ab} and
later extended in Ref.~\cite{bsa}, has evolved into the
``weak-coupling-plus'' (WCP) theory, developed and explored by
Rainer, Sauls, and Serene \cite{rs,sr,ser1}. This theory has
become a basic tool for the analysis of superfluid phases of
$^3$He, including the treatment of kinetic phenomena where the
quasiparticle picture \cite{lan} fails.

To describe these varied phenomena, Rainer et al.\ \cite{rs,sr,ser1}
have introduced about a dozen input parameters, composed of
density-dependent weighted angular averages of the normal-state
quasiparticle scattering amplitude.  In view of the proliferation
of phenomenological parameters in WCP theory, it is not well
suited for the task of disentangling the underlying reasons
for the NFL behavior of the {\it thermodynamic} properties
of liquid $^3$He and the attendant failings of the standard Landau-BCS
treatment.  It is our contention, to be supported in the following
analysis, that the prospects of resolving this issue are better,
if we proceed within the more transparent quasiparticle picture.
In this regard, we may call attention to earlier work in the
same spirit \cite{ckz}, in which the normal states of several
strongly interacting Fermi systems (notably 2D liquid $^3$He and
some heavy-fermion compounds) have been studied within an extended
quasiparticle picture.  The NFL behavior of these systems has been
successfully traced to deviations of the single-particle spectrum
from the usual Landau formula $\epsilon_{\fl}(p)=p_F(p-p_F)/M^*$,
which are in turn induced by the momentum dependence of the
effective interaction. Similarly, departures from the standard BCS
thermodynamic relations can appear if, in the single-particle spectrum
$E(p)=\sqrt{\epsilon^2(p)+\Delta^2(p)}$ of the superfluid state,
the gap function $\Delta(p)$ acquires a strong momentum dependence
driven by a characteristic momentum dependence of the pairing
interaction ${\cal V}$ (cf.\ Refs.~\cite{ap,lv,bp,pw,kkc}).

We note that the momentum dependence of $\Delta(p)$ is ignored
in the WCP theory.  Indeed, except in the arena of nucleonic
pairing \cite{kkc}, the momentum dependence of the superfluid
gap function and its empirical consequences are poorly understood.
Here we shall attempt to rectify this situation, in the context
of liquid $^3$He.

In Section 2, it is shown that the momentum dependence of
$\Delta(p)$ gives rise to deviations from the BCS thermodynamic
formulas for the zero-temperature gap and specific-heat jump.
Neglecting strong-coupling corrections, we derive a closed relation
between these deviations that is consistent with experimental
data \cite{todo,grey}. Strong-coupling corrections to the pairing
interaction ${\cal V}(p_1,p_2)$, vital for reproducing the phase
diagram of superfluid $^3$He \cite{ab,bsa}, are incorporated
 in Section 3.  We note, however, that such modifications of
${\cal V}$ cannot be the major factor in explaining the observed
departures from the two BCS thermodynamic relations.  For pressures
$P$ close to $P_{\rm max} \simeq 30$ bar, these corrections are responsible
 for only $\sim 1/3$ of the excess in the specific-heat
jump at the critical temperature, with lesser effects at lower
temperatures.

In Section 4, we return to a more detailed analysis of the
momentum dependence of the gap function, which becomes especially
strong when $\Delta(p)$ develops a pair of nodes located on either
side of the Fermi surface. We demonstrate that this phenomenon
occurs in the event that the pairing interaction ${\cal
V}(p_F,p_F)$ on the Fermi surface takes on a {\it positive} sign,
rather than remaining negative as conventionally assumed.  It is
argued that in liquid $^3$He the change of sign is caused by the
enormous enhancement of the repulsive component of the effective
interaction with increasing density at pressures approaching the
melting point. We find that the stronger the repulsive part of
${\cal V}$ is relative to its attractive component, the more
tightly this pair of nodes embraces the Fermi surface, and the
greater the anomalies in thermodynamic properties. Our conclusions
are summarized in Section 5, and their broader implications
examined.  It is conjectured that the condition ${\cal
V}(p_F,p_F)>0$ serves to promote $D$-pairing in a number of
heavy-fermion systems where antiferromagnetic fluctuations play no
significant part.

\section{Impact of the momentum dependence of the gap
function on the BCS relations}

\subsection{\it The B-state}

For superfluid $^3$He in the B-phase, non-BCS/NFL behavior
manifests itself in the first instance through experimentally
measured departures from the BCS relations
\beq
{\Delta(0)\over T_c}={\pi\over \gamma} \  , \qquad
 {C_s(T_c)-C_n(T_c)\over C_n(T_c)}={12\over 7\zeta(3)}=1.43
\label{rebcs}
\eeq
for the thermodynamic ratios $r_\Delta$ and $r_C$ appearing in
the leftmost members of the two equations.  Here $\ln\gamma=0.577$
is the Euler constant, and $\zeta(x)$, the Riemann zeta-function.

We now analyze the role of the momentum dependence of the gap
function $\Delta(p)$ in these deviations, neglecting
strong-coupling corrections for the time being.  In the B-state,
the BCS gap equation has the conventional form \beq
\Delta(p,T)=-\int{\cal V}(p,p_1){\tanh (E(p_1)/2T)\over
2E(p_1)}\Delta(p_1,T)d\upsilon_1\,,\label{bcs} \eeq where
$d\upsilon=p^2dp/ 2\pi^2$ and ${\cal V}(p,p_1)$, a real function,
denotes the effective $P$-wave pairing interaction. In BCS theory,
the quasiparticle energy $E(p)=\sqrt{\epsilon^2(p)+\Delta^2(p)}$
contains the single-particle energy in the normal state, given by
the FL formula $\epsilon(p)=p_F(p-p_F)/M^*$, where $M^*$ is the
effective mass.  The gap function $\Delta(p)$ is conveniently
written as a product \beq \Delta(p,T)\equiv \Delta (T) \psi(p) \,,
\label{shape0}
\eeq where $\Delta(T)\equiv \Delta(p=p_F,T)$ is the
magnitude of the gap, and the factor $\psi(p)$ represents its
shape, normalized by $\psi(p_F)=1$.  In what follows we neglect
the minor dependence of $\psi(p)$ on temperature, and assume
this function obeys the equation
\beq
\psi(p)=-\int{\cal
 V}(p,p_1){\tanh (\epsilon(p_1)/2T_c)\over
2\epsilon(p_1)}\psi(p_1)d\upsilon_1  \,, \label{shape}
\eeq
which is valid to a very good approximation.

Multiplication of Eq.~(\ref{bcs}) by the product $\psi(p)\tanh
(E(p)/2T)/2E(p_1)$, integration over momentum $p$, and other
simple manipulations lead to the identity
\beq
 \int\psi^2(p)\left[ {\tanh {\epsilon(p)\over 2T_c}\over
\epsilon(p)}-{\tanh{
\sqrt{\epsilon^2(p)+\Delta^2(T)\psi^2(p)}\over 2T}\over
\sqrt{\epsilon^2(p)+\Delta^2(T)\psi^2(p)}}\right]d\upsilon
=0\ .\label{rel1}
\eeq
In the BCS theory developed for conventional superconductors, the
pairing interaction is momentum-independent.  Hence $\psi \equiv
1$, and formula (\ref{rel1}) becomes
\beq
 \int\left[{\tanh
{\epsilon(p)\over 2T_c}\over \epsilon(p)}-{\tanh{
\sqrt{\epsilon^2(p)+\Delta^2_{\bcs}(T)}\over 2T}\over
\sqrt{\epsilon^2(p)+\Delta^2_{\bcs}(T)}}\right]d\upsilon =0
\,.\label{difbcs}
\eeq
Subtracting Eq.~(\ref{rel1}) from Eq.~(\ref{difbcs})
we arrive at
$$
\int{\tanh{ \sqrt{\epsilon^2(p)+\Delta^2_{\bcs}(T)}\over
2T}\over \sqrt{\epsilon^2(p)+\Delta^2_{\bcs}(T)}}\, d\upsilon =
\qquad\qquad\qquad\qquad
\qquad\qquad\qquad\qquad
$$
\beq
\qquad\qquad\qquad\qquad
=\int\left[(1-\psi^2(p))
{\tanh {\epsilon(p)\over 2T_c}\over
\epsilon(p)}+\psi^2(p){\tanh{
\sqrt{\epsilon^2(p)+\Delta^2(T)\psi^2(p)}\over 2T}\over
\sqrt{\epsilon^2(p)+\Delta^2(T)\psi^2(p)}}\right]\, d\upsilon  \,.
\label {difpr}
\eeq

Departures from the BCS relations (\ref{rebcs}) will be ascribed
to substantial momentum dependence of the gap function in
the region adjacent to the Fermi surface, where the integration
over $p$ can be replaced by an integration over the spectrum.
Thus we rewrite Eq.~(\ref{difpr}) as
$$
\int\limits_{-\infty}^{\infty}\left[{\tanh{
\sqrt{\epsilon^2+\Delta^2_{\bcs}(T)}\over 2T}\over
\sqrt{\epsilon^2+\Delta^2_{\bcs}(T)}}-{\tanh{
\sqrt{\epsilon^2+\Delta^2(T)}\over 2T}\over
\sqrt{\epsilon^2+\Delta^2(T)}}\right]\,d\epsilon =
\int\limits_{-\infty}^{\infty}
{(1-\psi^2(\epsilon))\left(\tanh{\epsilon\over 2T_c}
-\tanh {\epsilon\over 2T}\right)\over
\epsilon}\,d\epsilon
$$
\beq +\int\limits_{-\infty}^{\infty}\left[ (1-\psi^2(\epsilon))
{\tanh {\epsilon\over 2T}\over \epsilon} +\psi^2(\epsilon){\tanh{
\sqrt{\epsilon^2+\Delta^2(T)\psi^2(\epsilon)}\over 2T}\over
\sqrt{\epsilon^2+\Delta^2(T)\psi^2(\epsilon)}}-{\tanh{
\sqrt{\epsilon^2+\Delta^2(T)}\over 2T}\over
\sqrt{\epsilon^2+\Delta^2(T)}}\right]d\epsilon \,. \label{relpr}
\eeq

Now let us turn to the BCS relation (1) involving $\Delta(T=0)$.
Setting $T=0$ in Eq.~(\ref{relpr}), we have
 \beq
 \int\limits_{-\infty}^{\infty}\left[{1\over
\sqrt{\epsilon^2+\Delta^2_{\bcs}(0)}}-{1\over
\sqrt{\epsilon^2+\Delta^2(0)}}\right]d\epsilon= I_0(0)+I_1(0) \  ,
\label{relpr0} \eeq
or equivalently
 \beq
{\Delta^2(0)-\Delta_{\bcs}^2(0)\over\Delta^2(0)}=I_0(0)+I_1(0)  \ ,
\label{base0} \eeq
where
 \beq
I_0(0)=\int\limits_{-\infty}^{\infty}{(1-\psi^2(\epsilon) )(\tanh
 {|\epsilon|\over 2T_c} -1)\over|\epsilon|}d\epsilon  \  ,
\label{i00}
\eeq
and
\beq
I_1(0)=\int \limits_{-\infty}^{\infty}
\left[ {1\over |\epsilon|}- {1\over \sqrt{\epsilon^2+\Delta^2(0)}}
+\psi^2(\epsilon) \left(
{1\over\sqrt{\epsilon^2+\Delta^2(0)\psi^2(\epsilon)}} -{1\over
|\epsilon|} \right)\right]d\epsilon   \,. \label{i10}
 \eeq

The deviation of the jump in the specific heat at $T=T_c$ from its
BCS value (\ref{rebcs}) is due to the momentum dependence of the
gap function as $T\to T_c$.  One has $C=TdS/dT$ with $S= -\sum
\left[n_{{\bf p}}\ln n_{{\bf p}}+(1 -n_{{\bf p}}) \ln (1-n_{{\bf
p}})\right]$, so that the jump at $T _c$ is given by
\beq
C_s(T_c)-C_n(T_c) = -{1\over 2}\left({\partial \Delta^2(T\to T_c)
\over
\partial T}\right)_c \int \psi^2(p)\,n(p)(1-n(p))\,d\upsilon\,,
\label{spec1}
\eeq
where $n(p)=\left[1+\exp(\epsilon(p)/T)\right]^{-1}$.  As usual,
the corresponding BCS formula is obtained from this one by setting
$\psi\equiv 1$.  Upon neglecting a small contribution from the
integral containing the product
$(\psi^2(\epsilon)-1)n(\epsilon)(1-n(\epsilon))$, we find \beq
{C_s(T_c)-C_n(T_c)-\left[C_s(T_c)-C_n(T_c)\right]_{\bcs} \over
C_s(T_c)-C_n(T_c)}= {\left({\partial \Delta^2( T_c) \over \partial
T}\right)_c-\left({\partial \Delta^2_{\bcs}(T_c) \over \partial
T}\right)_c\over \left({\partial \Delta^2(T_c) \over \partial
T}\right)_c} \  . \label{rat1} \eeq

To proceed further, we take $T\to T_c$ in Eq.~(\ref{relpr}).
Manipulations similar to those resulting in Eq.~(\ref{base0}) then
lead us to the expression
$$
{7\zeta(3) \over 4\pi^2T_c^2} \left[ \Delta^2(T\to
T_c)-\Delta^2_{\bcs}(T\to T_c)\right] = {(T_c-T)\over 2T_c}
\int\limits_{-\infty}^{\infty}(\psi^2(\epsilon)-1 ){1\over
\cosh^2{\epsilon\over 2T_c}} d\epsilon$$
\beq + {1\over 2}
\Delta^2(T\to T_c) \int\limits_{-\infty}^ {\infty}
{(1-\psi^4(\epsilon))\over \epsilon^2} \left[ {\tanh{\epsilon\over
2T_c}\over\epsilon}- {1\over 2T_c|\epsilon|\cosh^2{\epsilon\over
2T_c}}\right]d\epsilon \,. \label{baset}
\eeq
In deriving this result, we have employed the identity \cite{lifpit}
\beq \int\limits_{-\infty}^{\infty} {dx\over 2x^2}
\left[{\tanh{x\over 2}\over x}-{1\over 2\cosh^2{x\over
2}}\right]={7\zeta(3) \over 4\pi^2}  \,.
\eeq

Upon eliminating $T_c-T$ with the aid of the BCS formula
\beq
\Delta^2_{\bcs}(T\to T_c)={8\pi^2\over 7\zeta(3)} T_c(T_c-T) \,,
\eeq
Eq.~(\ref{baset}) is rewritten as
\beq
 {\Delta^2(T\to
T_c)-\Delta^2_{\bcs}(T\to T_c)\over \Delta^2(T\to T_c)}
=I_0(T_c)+{4\pi^2\over 7\zeta(3)}I_1(T_c)
 \  , \label{reltc}
\eeq with \beq I_0(T_c)= {1\over 2T_c}
\int\limits_{-\infty}^{\infty}(\psi^2(\epsilon)-1 ){1\over
\cosh^2{\epsilon\over 2T_c}} d\epsilon  \  , \label{i0tc} \eeq and
\beq I_1(T_c) ={1\over 2}
T^2_c\int\limits_{-\infty}^{\infty}{(1-\psi^4(\epsilon))\over
\epsilon^2}\left[{\tanh{ \epsilon\over 2T_c}\over \epsilon}
-{1\over 2T_c|\epsilon|\cosh^2{ \epsilon\over 2T_c} }\right]
d\epsilon  \,. \label{i1tc} \eeq Now, since the l.h.s.\ of
Eq.~(\ref{reltc}) coincides with the r.h.s.\ of Eq.~(\ref{rat1}),
we can make the connection \beq
{C_s(T_c)-C_n(T_c)-\left[C_s(T_c)-C_n(T_c)\right]_{\bcs} \over
C_s(T_c)-C_n(T_c)}=I_0(T_c) +{4\pi^2\over 7\zeta(3)}I_1(T_c)  \,.
\label{rat2} \eeq
 It is seen that the integrands of the integrals
$I_0(0)$ and $I_0(T_c)$, as well as the last term in the integrand
of $I_1(T_c)$, become exponentially small for $|\epsilon|\gg T_c$.
At the same time, one sees from Eqs.~(\ref{i10}) and (\ref{i1tc})
that the leading terms in the integrands of $I_1(0)$ and
$I_1(T_c)$ fall off only as $\epsilon^{-1}$, at least in the
interval where $|\psi(\epsilon)|<1$.  Consequently, $I_0(0)$ and
$I_0(T)_c$ receive their overwhelming contributions from the
region $|\epsilon|\leq 2T_c$, implying that these integrals are
proportional to $T^2_c$.  On the other hand, both of the integrals
$I_1(0)$ and $I_1(T_c)$ contain an additional logarithmic factor
  $\sim \ln (\Omega_D/T_c)$ coming from the energy
region $T_c<|\epsilon|<\Omega_D$, where $\Omega_D$ is the Debye
frequency.
 We also note that for
$|\epsilon|>\Delta(\epsilon)\equiv \Delta(0)|\psi(\epsilon)|$, the
gap $\Delta$ can be neglected in the denominators of the integral
(\ref{i10}).  Accordingly, we can write
 \beq
I_1(0)/\Delta^2(0)\simeq I_1(T_c)/T^2_c={1\over 2}
\int\limits_{-\infty}^{\infty}{1-\psi^4(\epsilon)\over \epsilon^3}
\tanh{ \epsilon\over 2T_c}d\epsilon  \  . \label{finr} \eeq
Insertion of this relation into Eq.~(\ref{rat2}) leads finally to
  \beq
{C_s(T_c)-C_n(T_c)-\left[C_s(T_c)-C_n(T_c)\right]_{\bcs}\over
C_s(T_c)-C_n(T_c)}
 ={4\pi^2\over 7\zeta(3) }{T^2_c\over \Delta^2_{\bcs}(0)}
 \left({\Delta^2(0)-\Delta_{\bcs}^2(0)\over\Delta^2(0)}\right)  \,.
\label{finres} \eeq
Thus, we have derived a {\it closed relation}
between the departures of the two ratios $r_\Delta$ and $r_C$ from
their BCS values. Importantly, beyond the thermodynamic quantities
being connected, the relation {\it contains no input parameters}.

The existing experimental data only allow us to test this
relation for the B-state, and then only at $P=P_{\rm max}$, where
$\Delta(0)/\Delta_{\bcs}(0)\simeq 1.15$ \cite{todo}.  Upon
substituting this result into Eq.~(\ref{finres}) along with
$\Delta_{\bcs}(0)/T_c=1.76$, the calculated value of the
excess in the specific-heat jump at $T_c$ is found to
agree rather well with the experimental value of 30\% \cite{grey}.

\subsection{\it The A-state}

For the sake of clarity, we shall now distinguish gap functions in
the A and B phases by corresponding subscripts.  In the A-phase,
one has \cite{vw} $\Delta^2_{\aaa}({\bf p})\equiv \Delta^2(p)d^2({\bf
n}) $
with $d^2({\bf n})=(3/2) \sin^2\theta$, i.e., the gap function
depends not only on the absolute value of the momentum ${\bf p}$,
but also on its direction ${\bf n}$. In this case, the  BCS gap
equation (\ref{bcs}) takes the form
\beq
\Delta_{\aaa}(p)=-\int{\cal
V}(p,p_1)d^2({\bf n}_1) {\tanh (E({\bf p}_1)/2T)\over 2E({\bf
p}_1)} \Delta_{\aaa}(p_1){d\upsilon_1 d{\bf n}_1\over 4\pi}
\,,\label{bcsa} \eeq
with $E({\bf
p})=\sqrt{\epsilon^2(p)+\Delta^2_{\aaa}(p)d^2({\bf n})}$
and
$\Delta_{\aaa}(p) =\psi (p)\Delta_{\aaa}(T)$. The shape factor $\psi(p)$
again obeys Eq.~(\ref{shape}), independent of the structure of the
gap function.

Repeating the same manipulations as used to reach
Eq.~(\ref{rel1}), we now obtain \beq
\int\limits_{-\infty}^{\infty}\int\psi^2(\epsilon) \left[ {\tanh
{\epsilon\over 2T_c}\over \epsilon}-d^2({\bf n})
{\tanh{\sqrt{\epsilon^2+\Delta^2_{\aaa}(T)d^2({\bf n})\psi^2(\epsilon)}
\over 2T}\over \sqrt{\epsilon^2(p) +\Delta^2_{\aaa}(T)d^2({\bf n})
\psi^2(\epsilon)}}\right]{d\epsilon d{\bf n}\over 4\pi}\, =0\,.
\label{rela1} \eeq In the standard BCS theory, $\psi \equiv 1$ and
one obtains instead \beq
\int\limits_{-\infty}^{\infty}\int\left[{\tanh {\epsilon\over
2T_c}\over \epsilon}-d^2({\bf n}){\tanh{
\sqrt{\epsilon^2+\Delta^2_{\aaa}(T)d^2({\bf n})}\over 2T}\over
\sqrt{\epsilon^2+\Delta^2_{\aaa}(T)d^2({\bf n})}}\right] {d\epsilon
d{\bf n}\over 4\pi}\, =0  \,. \label{difbcsa} \eeq
 Upon approach
to the limit $T\to T_c$ where $\Delta(T)$ vanishes, the
denominator in Eq.~({\ref{difbcsa}) can be expanded about
$\Delta\equiv 0$; the integration over angles separates and is
freely performed. Then, comparing with Eq.(\ref{difbcs}), we come
to what is arguably the most problematic formula of the BCS theory
of superfluid $^3$He \cite{leg,vw}, namely
 \beq \left({\partial
\Delta^2_{\bbb}( T_c) \over \partial T}\right)_c =\kappa\left({\partial
\Delta^2_{\aaa}( T_c) \over \partial T}\right)_c\,, \label{dela} \eeq
with $\kappa=6/5$,
leaving no room on the phase diagram
for the A-phase -- a prediction in conflict with experiment.

Returning to Eq.~(\ref{rela1}) and considering temperatures close
to $T_c$, we see that repair of the BCS phase diagram of
superfluid $^3$He cannot be achieved by only
incorporation of the momentum dependence of the shape factor
$\psi(p)$, because of the separation of the integrations over the
direction and over the magnitude of the momentum $\bf p$. This
task is known to be the prerogative of strong-coupling corrections
\cite{ab} to the free energy of the superfluid state.

\section{Inclusion of strong-coupling corrections to the pairing interaction}

The origin and importance of strong-coupling corrections, proportional
to $\Delta^2(T)$ and reflecting the alteration of the pairing
interaction ${\cal V}$ in the superfluid state, have been elucidated
in many articles and books \cite{ab,bsa,leg,vw}.  In
strongly correlated Fermi systems, the rigorous evaluation of
these corrections, e.g.\ through the summation of parquet diagrams,
is impractical, so we treat their magnitude as a
phenomenological parameter.  Thus, for $T=T_c$ we write
\beq
{\left({\partial \Delta^2_{\bbb}( T_c)\over \partial T}\right)_c\over
\left({\partial \Delta^2_{\bcs}( T_c)\over \partial T}\right)_c}
=1+\delta_{\bbb}(T_c)\  ,  \qquad {\left({\partial \Delta^2_{\aaa}( T_c)
\over \partial T}\right)_c\over \left({\partial \Delta^2_{\bcs}(
T_c)\over \partial T}\right)_c} =\kappa^{-1}+\delta_{\aaa}(T_c) \,.
\eeq

A salient feature of the problem is the existence of a relation
between the quantities $\delta_{\aaa}(T_c)$ and $\delta_{\bbb}(T_c)$, stemming
from the structure of the order parameters in the B- and A-phases.
Correspondingly, $\delta_{\aaa}(T_c)\simeq 3\delta_{\bbb}(T_c)$, and the same
ratio holds at $T=0$ \cite{bsa}. Furthermore, the value of
$\delta_{\aaa}(T_c)$ must exceed $3(\kappa-1)/2\simeq 0.25$ in
order to outperform the factor $\kappa$ in Eq.~(\ref{dela}) and
protect the A-phase from extinction.  On the other hand,
at pressures $P\simeq P_{\rm max}$, the experimental values of
the A- and B-phase specific-heat jumps are the same within
10\%, indicating that $\delta_{\bbb}(T_c)\simeq 0.1$.  Based on
these considerations, the inclusion of strong-coupling corrections
alone (without accounting for the momentum structure of ${\cal V}$)
is incapable of explaining the observed departures from the BCS
relations (\ref{rebcs}). To wit: near $T_c$ these corrections
provide for only $1/3$ of the measured deviations, and according to
Ref.~\cite{bsa} their impact declines as $T \to 0$.
Thus we conclude that when one focuses on thermodynamic properties of
3D superfluid $^3$He within the quasiparticle picture, the shape factor
corrections are needed to resolve conflicts between the
experimental and theoretical values of the ratios $r_\Delta$
and $r_C$ appearing in Eq.~(\ref{rebcs}).

\section{Origin of nodes in the shape factor}

Ordinarily, in studies of superfluid Fermi liquids it is tacitly
supposed that the momentum dependence of the gap function is minor
and of negligible consequence.  However, the situation
dramatically changes, if the gap function develops a pair of nodes
lying on either side of the Fermi surface. This behavior plays a
decisive role in explaining the departures of the ratios
$r_\Delta$ and $r_C$ from their BCS values appearing in
Eq.~(\ref{rebcs}). The emergence of such a pair of nodes in the
shape function $\psi(p)$ does occur provided {\it the $P$-wave
pairing interaction ${\cal V}$ acquires the ``wrong'' (i.e.,
positive) sign  at the Fermi surface, ${\cal V}_F\equiv{\cal
V}(p_F,p_F)>0$}. At first glance, this condition appears to rule
out the existence of nontrivial solutions of the BCS gap equation
(\ref{bcs}).  But this is not the case.  Indeed, for some
plausible pairing interactions ${\cal V}$ used to describe
strongly correlated Fermi superfluids, it can happen that the sign
of ${\cal V}(p_1,p_2)$ is positive not only on the Fermi surface,
but {\it everywhere} in the $(p_1,p_2)$ plane, yet Eq.~(\ref{bcs})
nevertheless admits a nontrivial solution for the gap \cite{kkc}.
Such solutions can occur for an interaction ${\cal V}$ which,
viewed in coordinate space, possesses a strong inner repulsive
core and an outer attractive well.  The strength of the attraction
may be insufficient to compensate the repulsion in forming the
pairing matrix elements in momentum space, but still strong enough
to create a Cooper-pair instability.

In what follows we employ the separation method that has been
developed for efficient analysis and solution of gap equations
\cite{kkc}.  The defining step of this procedure consists in
decomposing the interaction ${\cal V}(p_1,p_2)$ into a separable
part and a remainder $W(p_1,p_2)$ that vanishes when either
argument is at the Fermi surface:
 \beq {\cal V}(p_1,p_2) = {\cal
V}_F\varphi(p_1)\varphi(p_2) + W(p_1,p_2)   \,. \label{decom} \eeq
The choice $\varphi(p)={\cal V}(p,p_F)/{\cal V}_F$ meets the
required condition $W(p_F,p)=W(p,p_F)=0$ for all $p$.  Upon
inserting the decomposition (\ref{decom}) into Eq.~(\ref{bcs}) one
finds \cite{kkc} \beq \psi(p)  = \varphi(p)-\int
W(p,p_1){\tanh{\epsilon(p_1)\over 2T_c}\over
2\epsilon(p_1)}\psi(p_1)d\upsilon_1 \  , \label{eqshap} \eeq and
\beq 1=-{\cal V}_F\int \varphi(p)\psi(p){\tanh{\epsilon(p)\over
2T_c}\over 2\epsilon(p)} d\upsilon \  . \label{emag} \eeq Since
$\psi(p_F)=\varphi(p_F)=1$, the sign of the leading logarithmic
part $\sim \ln (1/T_c)$ of the r.h.s.\ of Eq.~(\ref{emag}), being
negative due to ${\cal V}_F>0$, is opposite to that of the l.h.s.
However, if the shape factor $\psi(p)$ changes its sign (see
below), and this node occurs so close to the Fermi surface that
the remaining part of the integral on the r.h.s.\ of
Eq.~(\ref{emag}) becomes positive and overweighs unity, then
nontrivial solutions of the gap equation exist \cite{kkc}.  The
number of nodes of the function $\psi(p)$ depends on the structure
of the pairing potential ${\cal V}$, and in particular, on the
relative strength of its repulsive and attractive components,
which governs the sign and value of the key parameter ${\cal
V}_F$.

To illustrate and affirm these assertions, we choose a pairing
interaction ${\cal V}({\bf p}_1, {\bf p}_2)$ of the form \beq
{\cal V}({\bf p}_1, {\bf p}_2)= { V_r\over ({\bf p}_1-{\bf
p}_2)^2+\beta^2_r}-{ V_a\over ({\bf p}_1-{\bf p}_2)^2+\beta^2_a}\,
.\label{yuk} \eeq
 Such a model pairing potential mimics the
realistic pairing interaction in superfluid $^3$He, which features
a repulsive core and a long-range attractive part \cite{pit,kw}.
To treat $P$-pairing, we evaluate the first harmonic of this
interaction over the angle between ${\bf p}_1$ and ${\bf p}_2$
and employ the result in Eq.~(\ref{eqshap}).

First we elucidate the emergence of the pair of the nodes in the
shape factor $\psi(p)$ in the case of ${\cal V}_F>0$.  Fig.~1
shows results from solution of Eq.~(\ref{eqshap}) with two sets of
input parameters: (i) $\beta_r=p_F$, $\beta_a=0.1p_F$,
$V_aN(0)=-3.0$, and $V_rN(0)=45$, where $N(0)=p_FM^*/\pi^2$ is the
density of states; and (ii) $\beta_r=p_F$, $\beta_a=0.05p_F$,
$V_aN(0)=-5.5$, and $V_rN(0)=90$.  In both of these cases, the
value of the dimensionless coupling parameter $N(0){\cal
V}_F\simeq 2.0$ is positive and quite large.

%%%%%%%%%%%%%%%%%%%%%%%%%%%%%%%%%%%%%%%%%%%%%%%%%%%%%%%%%%%%%%%%%
%%%%%                Figure: shape
%%%%%%%%%%%%%%%%%%%%%%%%%%%%%%%%%%%%%%%%%%%%%%%%%%%%%%%%%%%%%%%%%
\begin{figure}[t]
\begin{center}
\includegraphics[width=0.65\linewidth,height=0.8\linewidth]{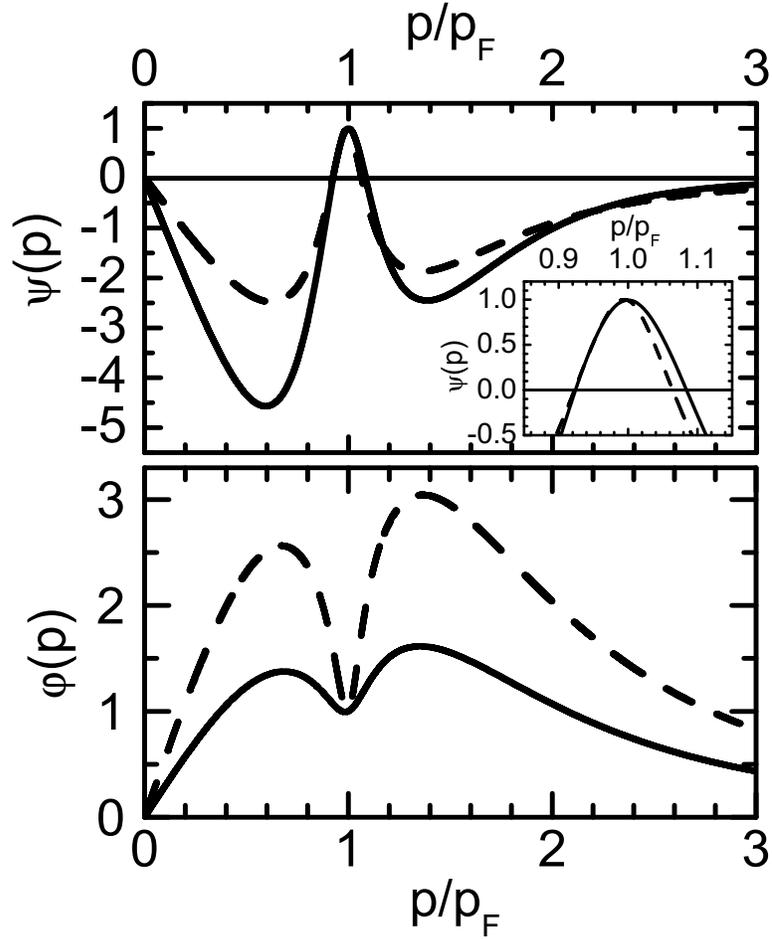}
\end{center}
\hskip 1.5 cm
\begin{minipage}[c]{0.8\linewidth}
\caption{Shape factor $\psi(p)$ (top panel) and interaction
profile $\varphi(p)$ (bottom panel) versus $p/p_F$. The solid
(dashed) line corresponds to the case $\beta_a=0.1\,p_F$, ${\cal
V}_FN(0)=2.0$ ($\beta_a=0.05\,p_F$, ${\cal V}_FN(0)=2.5$). In both
cases $T_c\simeq 10^{-3}\epsilon^0_F$. The inset depicts the
behavior of the shape factor in the vicinity of the
Fermi surface.}
\end{minipage}
\label{fig:shape}
\end{figure}
%%%%%%%%%%%%%%%%%%%%%%%%%%%%%%%%%%%%%%%%%%%%%%%%%%%%%%%%%%%%%%%%

The lower panel of Fig.~1 shows $\varphi(p)\equiv {\cal
V}(p,p_F)/{\cal V}(p_F,p_F) $, the interaction profile in the
$P$-pairing channel. This function has the shape of a wide hump,
with a narrow dip at $p=p_F$. The $P$-wave shape factor $\psi(p)$,
which vanishes for $p\to 0$, is shown in the upper panel.  It
exhibits a narrow, sharp peak at the Fermi surface that creates
two nodes, one on either side of the Fermi surface and lying very
close to $p_F$.

%%%%%%%%%%%%%%%%%%%%%%%%%%%%%%%%%%%%%%%%%%%%%%%%%%%%%%%%%%%%%%%%%
%%%%%                Figure: nodes
%%%%%%%%%%%%%%%%%%%%%%%%%%%%%%%%%%%%%%%%%%%%%%%%%%%%%%%%%%%%%%%%%
\begin{figure}[t]
\begin{center}
\includegraphics[width=0.75\linewidth,height=0.6\linewidth]{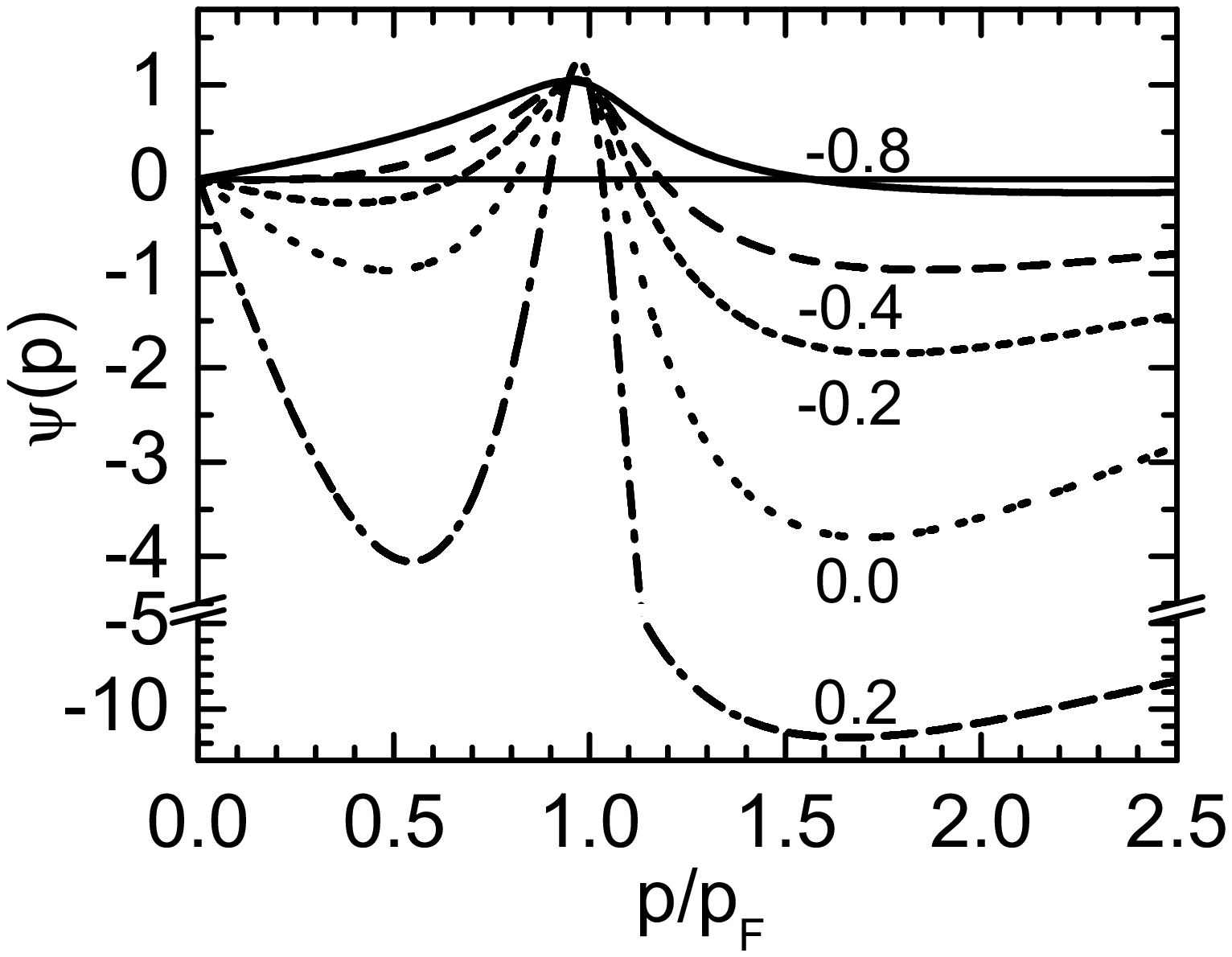}
\end{center}
\hskip 1.5 cm
\begin{minipage}[c]{0.8\linewidth}
\caption{$P$-pairing shape factor $\psi(p)$ as a function of
$p/p_F$. The interaction form (\ref{yuk}) is assumed, with $
V_aN(0)=1$, $\beta_a/p_F=0.1$, $\beta_r/p_F=2$, and five different
values of the strength parameter $V_r$ of the repulsive component.
The curves are labelled with the corresponding values of the
$P$-wave coupling parameter $v_F\equiv{\cal V}_FN(0)$.}
\end{minipage}
\label{fig:nodes}
\end{figure}
%%%%%%%%%%%%%%%%%%%%%%%%%%%%%%%%%%%%%%%%%%%%%%%%%%%%%%%%%%%%%%%%

Now we address the question: how does the number of nodes in the shape
factor $\psi(p)$ change as ${\cal V}_F$ is varied while holding
$V_a$ fixed?  Evidently, if the repulsion is  weak, then the shape
factor $\psi(p)$ has no nodes at all.  However, if the repulsion
begins to rise, such that ${\cal V}_F$ attains a critical value
${\cal V}_{1F}$, the first node of $\psi(p)$ emerges at some
momentum $p_0 > p_F$.  Whether
the sign of ${\cal V}_{1F}$ remains negative or already becomes
positive will depend on the particular form of the pairing interaction
and on the input parameters that specify it.  With further increase
of ${\cal V}_F$, the nodal location $p_0$ moves toward the
Fermi surface from the outside.
 However, the destination $p_0=p_F$ turns out to be
unattainable \cite{kkc}.

Meanwhile, as ${\cal V}_F$ continues to increase and reaches a
critical value ${\cal V}_{2F}$, a second node of $\psi(p)$ emerges
at some momentum value below $p_F$ and moves rapidly toward the
Fermi surface from the inside.  As we shall see, the two nodes
crowd the Fermi surface more and more tightly as ${\cal V}$
rises higher and higher above the threshold ${\cal V}_{2F}$.

Fig.~2 presents results from numerical calculations of the shape
factor $\psi(p)$, based again on the the interaction ${\cal V}$
defined in Eq.~(\ref{yuk}).  However, we now choose $V_aN(0)=-1.0$
and illustrate the sensitivity of the results to the potential
parameters.  The choice $\beta_a=0.1p_F$ is maintained, but
$\beta_r$ is increased to $2p_F$.  From this figure, we may infer
that the first node of $\psi(p)$ emerges at some large momentum $p
\sim \infty$ when $v_F \equiv {\cal V}_{1F}N(0)\simeq -1.0$. Also,
it is important to note that in contrast to the situation found in
Fig.~1, the second node of the shape factor now comes on the scene
while ${\cal V}_F$ is still negative, with a value ${\cal
V}_{2F}N(0)\simeq -0.3$.  Thus, we conclude that the sign of the
critical value of the parameter ${\cal V}_F$ for emergence of the
second node of $\psi(p)$ depends nontrivially on the specific
choice of input parameters.
%%%%%%%%%%%%%%%%%%%%%%%%%%%%%%%%%%%%%%%%%%%%%%%%%%%%%%%%%%%%%%%%%
%%%%%                Figure: two
%%%%%%%%%%%%%%%%%%%%%%%%%%%%%%%%%%%%%%%%%%%%%%%%%%%%%%%%%%%%%%%%%
\begin{figure}[t]
\begin{center}
\includegraphics[width=0.5\linewidth,height=0.8\linewidth]{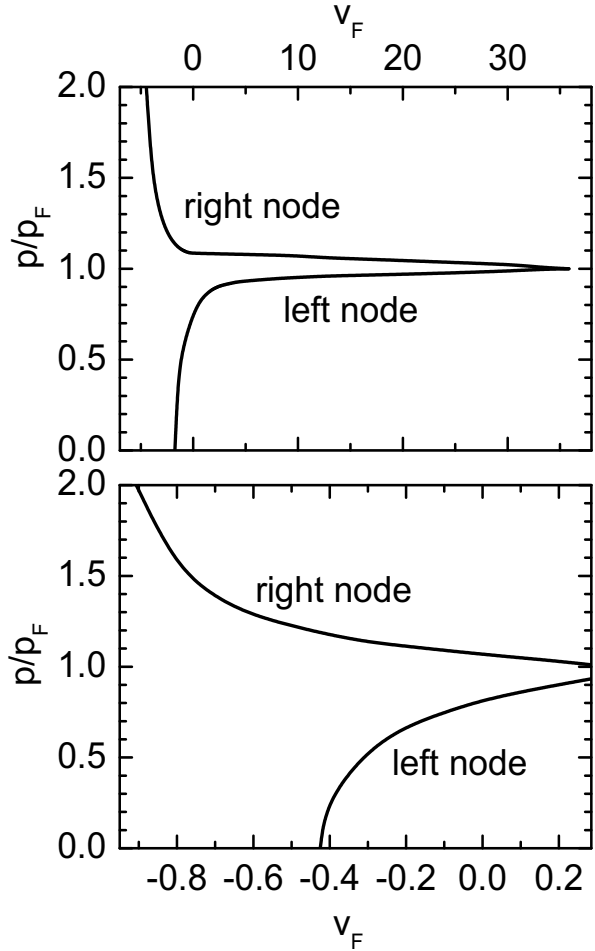}
\end{center}
\hskip 1.5 cm
\begin{minipage}[c]{0.8\linewidth}
\caption{Trajectories of the pair of nodes that appear in the
$P$-pairing shape function $\psi(p)$ of Fig.~1 (top panel) and
Fig.~2 (bottom panel), versus the $P$-wave coupling parameter
$v_F$.}
\end{minipage}
\label{fig:two}
\end{figure}
%%%%%%%%%%%%%%%%%%%%%%%%%%%%%%%%%%%%%%%%%%%%%%%%%%%%%%%%%%%%%%%%

 In Fig.~3, we display the trajectories of the nodes of the shape
factor $\psi(p)$ under variation of the key parameter ${\cal
V}_F$,
%employing the same set of the rest parameters, as in Figs.~1 and 2.
with the same parameter setups as in Figs.~1 and 2. The two nodes
rapidly approach the Fermi surface from opposite sides as $v_F=
{\cal V}_F N(0)$ tends to toward a critical value $v_{Fc}$.  (Once
again,
as seen,
 this critical value  depend crucially on the specifics of
the input potential parameters.)

%%%%%%%%%%%%%%%%%%%%%%%%%%%%%%%%%%%%%%%%%%%%%%%%%%%%%%%%%%%%%%%%%
%%%%%                Figure: phd
%%%%%%%%%%%%%%%%%%%%%%%%%%%%%%%%%%%%%%%%%%%%%%%%%%%%%%%%%%%%%%%%%
\begin{figure}[t]
\begin{center}
\includegraphics[width=0.65\linewidth,height=0.55\linewidth]{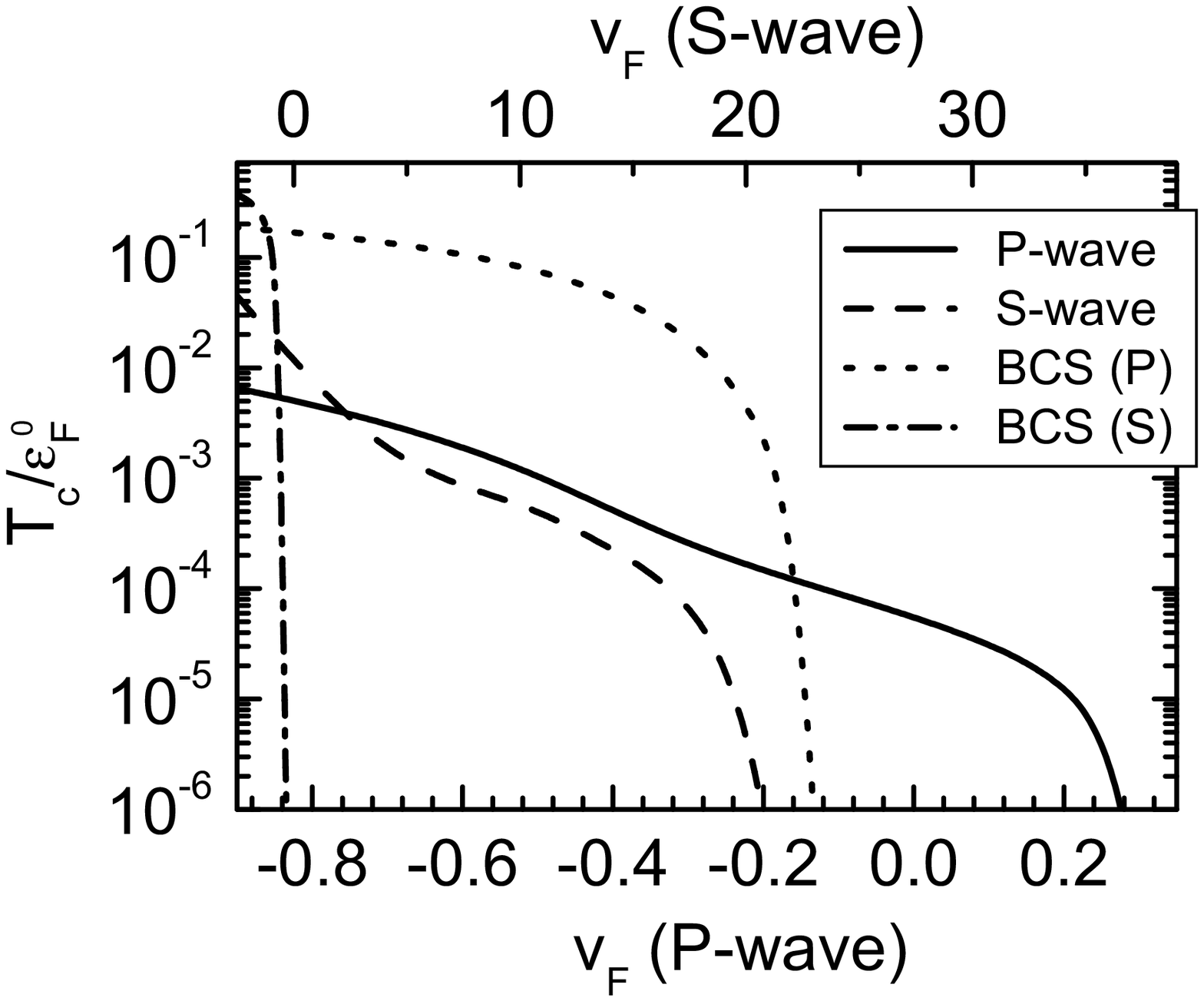}
\end{center}
\hskip 1.5 cm
\begin{minipage}[c]{0.8\linewidth}
\caption{Critical temperature $T_c$ (in units of
$\varepsilon_F^0=p_F^2/2M$), plotted against $P$-wave (bottom axis) and
$S$-wave (top axis) coupling parameters $v_F \equiv {\cal V}_FN(0)$,
for a model interaction of form (\ref{yuk}).  The solid
curve corresponds to $P$-pairing, and the dashed curve, to
$S$-pairing.  For comparison we show the BCS estimate (\ref{bcstc})
of $T_c$ for $P$-pairing (dotted curve) and for $S$-pairing
(dot-dashed curve).}
\end{minipage}
\label{fig:phd}
\end{figure}
%%%%%%%%%%%%%%%%%%%%%%%%%%%%%%%%%%%%%%%%%%%%%%%%%%%%%%%%%%%%%%%%

\section{Discussion}
A complete theory of superfluid $^3$He must give a quantitative
description of its kinetic phenomena as well as its thermodynamic
properties.  Necessarily, then, such a theory takes explicit
account of the damping of single-particle excitations -- as, for
example in the weak-coupling-plus (WCP) theory
\cite{bsa,rs,sr,ser1}. By virtue of the scope of WCP theory, its
phenomenological character entails the introduction of a large
number of adjustable parameters, which inevitably complicates the
interpretation of new experimental results.  In contrast to this
approach, our concentration on the thermodynamics of superfluid
$^3$He has allowed us to retain the conceptually simpler Landau
picture \cite{lan}.  In essence, this picture stems from the basic
statement, or assumption, that the ground-state energy and other
thermodynamic quantities are functionals of the quasiparticle
momentum distribution $n(p)$. Seemingly innocuous at first sight,
the Landau assumption leads rather directly to the determination
of this distribution function, which can be expressed in the same
form as the Fermi-Dirac momentum distribution,
$n(p)=[1+\exp(\epsilon(p)/T)]^{-1}$.  There is an important
distinction, however, namely that the Landau quasiparticle
spectrum $\epsilon_{\fl}(p)=p_F(p-p_F)/M^*$ differs from the
single-particle spectrum of the ideal Fermi gas, since it involves
an effective mass $M^*$ different from the bare mass $M$. In fact,
this formula for $\epsilon_{\fl}$ is the most vulnerable
element of Landau theory as traditionally practiced. The spectrum
$\epsilon_{\fl}(p)$ ceases to be meaningful close to the quantum
critical point (QCP) where the effective mass $M^*$ diverges. To
rectify the theory, one must take into account contributions to
$\epsilon(p)$ from terms $\sim (p-p_F)^3$, as established in
Ref.~\cite{ckz}.  This extension of the Landau quasiparticle
picture alters the standard Fermi-liquid formulas, with the
implication that in the vicinity of the QCP, (so-called)
non-Fermi-liquid behavior is predicted within the Fermi-liquid
approach itself.

A parallel situation may arise in superfluid Fermi liquids.
Non-BCS behavior can be deduced within the standard BCS approach
provided a well-pronounced momentum dependence of the gap function
$\Delta(p)$, driven by the momentum dependence of the pairing
interaction, is properly incorporated.  In liquid $^3$He, the bare
atom-atom interaction, as modelled in Ref.~\cite{aziz} by a local
two-body potential function in coordinate space, contains a huge
repulsive core of radius $r_c\simeq 2.5\,\AA$, surrounded by an
attractive, long-range (power-law) van der Waals component. The
properties of this in-vacuum interaction are presumably mirrored
by an in-medium pairing interaction ${\cal V}$ consisting of a
local repulsive component at short distance and a longer-range
attractive term \cite{bp,pit}.  From the perspective offered and
supported in this paper, the existence of $P$-wave superfluidity
in liquid $^3$He implies the presence of an attractive component
$V_a$ of the effective pairing interaction ${\cal V}$ strong
enough to outperform the repulsion at distances slightly exceeding
the average distance $r_0$ between particles.  On the other hand,
at large pressures, the repulsive part $V_r$ of the effective
interaction increases rapidly with the density $\rho$.  Indirect
evidence for the latter inference is seen in the fast growth of
the sound velocity at pressures $P\simeq P_{\rm max}$.

Let us suppose that the parameter ${\cal V}(p_F,p_F)$ has already
turned positive at pressures close to $P_{\rm max}$, resulting
in the appearance of two nodes in the shape factor
$\psi(p)$ of the gap function that hold the Fermi surface
in a ``vice-like grip.''  Furthermore, to facilitate the analysis,
let us assume that these nodes are situated symmetrically
with respect to the Fermi surface
on either side of it,
so that their locations may be specified by a single
parameter. The plots of $\psi(p)$ obtained for models based on the
pairing interaction (\ref{yuk}) indicate that this assumption is
not unreasonable (see Figs.~1 and 2).  It becomes more reasonable
when long-wavelength spin fluctuations, analogous to the mediating
phonons of conventional superconductors, dominate in the
attractive part of ${\cal V}$, since in this case the induced
pairing interaction is symmetrical with respect to departures of
the quasiparticle momenta from the Fermi surface.

Numerical analysis shows that the dominant contributions to the
integral $I_1(0)$ of Eq.~(\ref{i10}) come from the interval
$[-\epsilon_0,\epsilon_0]$.  Upon neglecting the contributions
from the remainder of momentum space and introducing a new
integration variable $x=\epsilon/\epsilon_0$, the integral
(\ref{i10}) takes the form
\beq
I_1(0)=(\Delta^2(0)/\epsilon_0^2)K(0)\ ,
\label{ik}
\eeq
where
\beq K(0)=2\int\limits_0^1 {1-\psi^4(x)\over
x\sqrt{x^2+\Delta^2(0)/\epsilon^2_0}\left(x+\sqrt{x^2+\Delta^2(0)
/\epsilon^2_0}\right)}dx
\ . \label{i11}
\eeq
It should be emphasized that deviations of the ratios $r_{\Delta}$
and $r_C$ from their BCS values proportional to the ratio
$\Delta^2(0)/\epsilon_0^2$ will depend crucially on the
structure the shape factor $\psi(x)$ in the interval $0<x<1$.
To support this statement, we calculate the value of the integral
(\ref{i11}) for two phenomenological shape factors: (i) $\psi(x)=1-x^2$
and (ii) $\psi(x)=(1-x^2)^{\alpha}$.  In the latter case we take
$\alpha=1.6$, a value that provides an adequate fit of $\psi(x)$
as given in Fig.~1.  Calculation of the integral in (\ref{i11})
yields $K(0)=6$ in the first case and $K(0)=50$ in the second.

We may attempt to narrow the uncertainty in the value of the
parameter $\epsilon_0$, based on the observation that
in the normal state at pressures $P$ close to the melting point,
the experimental spin susceptibility $\chi(T)$ coincides with
the Curie susceptibility $\chi_C(T)=\rho/T$ at $T\geq 0.4$ K.
This finding implies that at $P\simeq P_{\rm max}$, the
bandwidth $\Omega = \epsilon(p_F) - \epsilon(p=0)$ may be
estimated as 0.2--0.3 K.  We then suggest that for
$P\simeq P_{\rm max}$, the value of $\epsilon_0$, which must
be significantly less than $\Omega$, lies in the interval
0.05--0.1 K.  Inserting this estimate into Eq.~(\ref{ik}),
along with the two values obtained for $K(0)$,
one obtains a range of values for
the deviation of  $r_\Delta$ from the BCS value in the interval
1\%-50\%.

In Fig.~4 we present results from calculations of the critical
temperature $T_c$ for $P$-pairing and $S$-pairing as a functions
of the coupling parameter $v_F = {\cal V}_FN(0)$. Once again, a
pairing interaction of form (\ref{yuk}) is assumed
to have the
same set of parameters as in Fig.~2.
Here we also take the opportunity to highlight the problematic
nature of the standard BCS formula commonly used to estimate the
critical temperature for pair condensation in fermionic systems,
\beq
T_c={2\gamma\over \pi}\Omega_D \exp\left(-{2\over
\lambda N(0)}\right) \  ,
\label{bcstc}
\eeq
wherein
$N(0)=p_FM^*/\pi^2$ is the density of states, $\Omega_D$ is the
Debye frequency, and $\lambda$ is the coupling constant, usually
identified with $-{\cal V}_F$. Fig.~4 compares critical
temperatures obtained with this formula (curves labelled BCS(P)
and BCS(S)) with those determined by direct solution of the gap
equation taking full account of the shape dependence of the gap
function. Evidently, the estimate (\ref{bcstc}) is irrelevant if
${\cal V}_F>0$, so meaningful comparisons can only be made for
$v_F > 0$. It is seen that the formula (\ref{bcstc}) already
begins to fail well before ${\cal V}_F$ changes its sign.  In
general, then, the domain where superfluidity exists is wider than
that where ${\cal V}_F$ is negative, a fact long appreciated
within the theory of nucleonic superfluids (see
Refs.~\cite{kc,ccdk,kkc} and works cited therein). In light of
this conclusion, the findings of Kuchenhoff and W\"olfle \cite{kw}
on the occurrence of superconductivity in the dilute electron gas
bear re-examination, since these authors determined the extent of
the superconducting phase from the condition ${\cal V}_F<0$.
Referring once more to Fig.~4, we may also note that $P$-pairing
wins the energetic contest with $S$-pairing when the key parameter
${\cal V}_F$ takes on positive values.

A similar situation may exist for unconventional $D$-pairing in
heavy-fermion metals.  A conspicuous example is CeCoIn$_5$, for
which the antiferromagnetic state lies well above the ground
paramagnetic state, implying that antiferromagnetic fluctuations
are irrelevant to pairing in the ground state.  In principle,
there are several sources for attraction in the pairing
interaction ${\cal V}$; these include optical phonons and density
fluctuations associated with the proximity to a critical density
where electron liquid ceases to be homogeneous \cite{bz}.  We
argue that the attractive part of ${\cal V}$ changes
insignificantly in switching from $S$- to $D$-pairing, while the
repulsive part drops substantially.  Thus, in certain
heavy-fermion metals, unconventional $D$-pairing may win the
contest with conventional $S$-pairing if there is sufficient
weakening of the repulsion in the $D$-wave channel relative to
that in the $S$-wave channel.

In conclusion, we have analyzed the anomalous behavior of certain
thermodynamic properties of superfluid phases of $^3$He in the
framework of the Landau quasiparticle picture.  Principally, this
behavior consists of departures of experimental results from two
famous BCS relations, one quantifying the ratio of the pairing gap
at zero temperature to the critical temperature, and the other,
the ratio, at this temperature, of the difference between normal
and superfluid specific heats to the normal specific heat,  We
have proposed and demonstrated that, within the quasiparticle picture,
these discrepancies may be traced to the emergence of a pair
nodes in the gap function lying close to and on
opposite sides of the Fermi surface.  We have found that a
quantitative explanation of the anomalies requires only two
phenomenological parameters, one specifying the location of the
nodes of the gap function, and the other setting the scale of
strong-coupling corrections.  Finally, it should be noted
that our phenomenological theory has little in common with
the (likewise phenomenological) weak-coupling-plus theory
of Refs.~\cite{bsa,rs,sr,ser1}, in which the momentum
dependence of the gap function is completely ignored.

\end{document}